\documentclass[iop,apj,twoside]{emulateapj}
\usepackage{amsmath,amssymb,amstext}
\usepackage[breaklinks,colorlinks,citecolor=blue,linkcolor=magenta]{hyperref} 

\usepackage[all]{hypcap}
\usepackage{booktabs}
\usepackage{natbib}
\usepackage{xspace}

\newcommand{\kms}[0]{km~s$^{-1}$\xspace}
\newcommand{\msyr}[0]{M$_{\Sun}$~yr$^{-1}$\xspace}
\newcommand{\msyrpc}[0]{M$_{\Sun}$~yr$^{-1}$~pc$^{-2}$\xspace}

\newcommand{\sfrd}[0]{$\Sigma_{\rm SFR}$\xspace}
\defcitealias{Tanner1}{Paper I}

\shorttitle{Scaling Relations of Galactic Winds}
\shortauthors{Tanner, Cecil, \& Heitsch}

\begin{document}

\title{Scaling Relations of Starburst-Driven Galactic Winds}

\author{Ryan Tanner, Gerald Cecil, and Fabian Heitsch}
\affil{University of North Carolina at Chapel Hill\\
Chapel Hill, NC 27599-3255\\
rjtanner@physics.unc.edu}

\begin{abstract}
Using synthetic absorption lines generated from 3D hydro-dynamical simulations we explore how the velocity of a starburst-driven galactic wind correlates with the star formation rate (SFR) and SFR density. 
We find strong correlations until the scaling relations flatten abruptly at a point set by the mass loading of the starburst. 
Below this point the scaling relation depends on the temperature regime being probed by the absorption line, not on the mass loading. 
The exact scaling relation depends on whether the maximum or mean velocity of the absorption line is used. 
We find that the outflow velocity of neutral gas is four to five times lower than the average velocity of the hottest gas, the difference increasing with gas ionization. 
Thus, absorption lines of neutral or low ionized gas will underestimate the outflow velocity of hot gas, severely underestimating outflow energetics.
\end{abstract}

\section{Introduction\label{sec:intro}}
Observations beginning in the 1990s established galactic winds as ubiquitous phenomena associated with star-forming galaxies \citep{1993ASSL..188..455H,1995PASA...12..190B,1997PASP..109.1298D,2000ApJS..129..493H}. 
These observations focused on optical emission lines images and spectroscopy \citep{1993ASSL..188..455H}. 
Optical imagery established the physical morphology of galactic winds and spectroscopy provided the kinematics and warm plasma diagnostics. 
While emission traced the interaction of the warm ISM with the hot wind, absorption lines probed the interaction between warm and cold gas and the hot wind \citep{2000ApJS..129..493H}. 
X-ray emission, first observed in M82 \citep{1984ApJ...286..144W}, would also become important for identifying galactic outflows and measuring wind energetics \citep{1988ApJ...330..672F,1990ApJ...355..442F,1993ASSL..188..455H,1995ApJ...448...98H}. While some studies of galactic winds focused on X-ray emission \citep{StricklandStevens,2009ApJ...697.2030S}, \citet{1995PASA...12..190B} predicted that multi-band observations of galactic winds would become standard in characterizing galactic winds, and \citet{2005ARA&A..43..769V} have shown that subsequent multi-band studies are important in characterizing the galactic wind.

More recent observations \citep{2012ApJ...760..127M,2014A&A...568A..14A,2014ApJ...794..156R,2015ApJ...811..149C,2015ApJ...809..147H,2016ApJ...822....9H} show that galactic winds are ubiquitous for star forming galaxies. 
Galactic winds are detected in 45\% \citep{2012ApJ...760..127M}, 74\% \citep{2015ApJ...811..149C}, 66\% (89\% for face on, 45\% for edge on galaxies) \citep{2014ApJ...794..156R}, and 90\% \citep{2015ApJ...809..147H,2016ApJ...822....9H} of star forming galaxies surveyed. 
Outflow kinematics are typically measured using UV absorption lines such as: Mg II and Fe II \citep{2014ApJ...794..156R}, Si II, Si III, Si IV and O I \citep{2015ApJ...811..149C,2016MNRAS.457.1257H,2015ApJ...809..147H,2016ApJ...822....9H}, Na D \citep{2000ApJS..129..493H,2005ApJ...621..227M}, H$\alpha$ and O III \citep{2016A&A...588A..41C}, and C II and N II \citep{2015ApJ...809..147H,2016ApJ...822....9H}.

\citet{2000ApJS..129..493H} found that starburst galaxies whose Na D absorption line is dominated by the ISM, typically exhibited outflow velocities of $> 100$ \kms, with maximum velocities ranging from $300-700$ \kms. 
They were able to map outflow gas up to 10 kpc from the galactic center. 
They concluded that dense clouds in the ISM were being disrupted by the galactic wind, and that the ablated gas was being accelerated up to the terminal wind velocity.

\citet{2005ApJ...621..227M} investigated the relationship between outflow velocities, as measured by the Na D lines, and the SFR. 
She found that the maximum wind velocity correlates as SFR$^{.35}$, and that stellar luminosity suffices to accelerate cool outflows to the terminal velocity. 
Martin noted that the covering fraction of the cold gas is not complete, which indicates that it is not a continuous fluid but is broken into clouds or shells.

\citet{2014ApJ...794..156R} extended previous work using Mg II and Fe II absorption lines to find that outflows are detected for all $M_\star$, SFR and SFR density (\sfrd) studied. 
Interestingly they found no evidence of a minimum threshold for \sfrd. 
This indicates that galactic winds can still form in galaxies with extremely low \sfrd. 
Although outflows were detected for all parameter ranges, a correlation was only found between outflow velocity and $M_\star$. 
These findings are both consistent with and conflict with previous work \citep{2009ApJ...692..187W,2010AJ....140..445C,2011ApJ...730....5H,2012ApJ...760..127M}.

Conversely \citet{2015ApJ...811..149C} found correlations between $M_\star$ and SFR, but not with \sfrd, using Si II absorption lines. 
They found a weak correlation between SFR and maximum velocity, but a slightly stronger correlation between SFR and the velocity as measured by the line center. 
In agreement with \citet{2014ApJ...794..156R}, \citet{2015ApJ...811..149C} found that there is no minimum \sfrd at which outflows form.

\citet{2016A&A...588A..41C} studied a large sample of galaxies and found that the outflow velocity correlates with both the SFR and the specific SFR, but only for a SFR $> 1$~\msyr. 

\citet{2015ApJ...809..147H} and \citet{2016ApJ...822....9H} found that the outflow velocity correlates strongly with SFR and \sfrd but weakly with galactic stellar mass. 
They also also noted that at sufficiently high \sfrd the correlation flattens out. 

In this paper we resolve these differences and explain why some surveys of starburst galaxies have found correlations between the outflow velocity of starburst-driven winds and both SFR and \sfrd, while others have not. 
We also address why even among studies of star forming galaxies that have found a correlation, there is no agreement on the measured scaling relation. 
We use 3D hydro models of nuclear starbursts to study how the outflow velocities of galactic winds scale with SFR and \sfrd. 
In Section \ref{sec:setup} we explain our setup and model parameters. 
In Section \ref{sec:lines} we describe how we generate synthetic absorption profiles for Si ions to use as tracers of outflow velocity at different gas temperatures. 
In Section \ref{sec:res} we describe how the outflow velocity correlates with SFR and \sfrd, followed with a discussion in Section \ref{sec:dis} of how our results relate to observational results. 

%Setup
\section{Setup}\label{sec:setup}
Our models use the setup of \citet[][hereafter \citetalias{Tanner1}]{Tanner1} with a few modifications described below. As in \citetalias{Tanner1} we simulate a nuclear starburst inside a box 1000 pc on a side. The stellar gravitational potential and associated parameters are set to correspond to an M82 sized galaxy. All models are run for 1.5 Myr.

\subsection{Static Mesh Refinement}\label{sec:setup:smr}
\begin{figure}[!ht]
\centering
\includegraphics[width=0.5\textwidth]{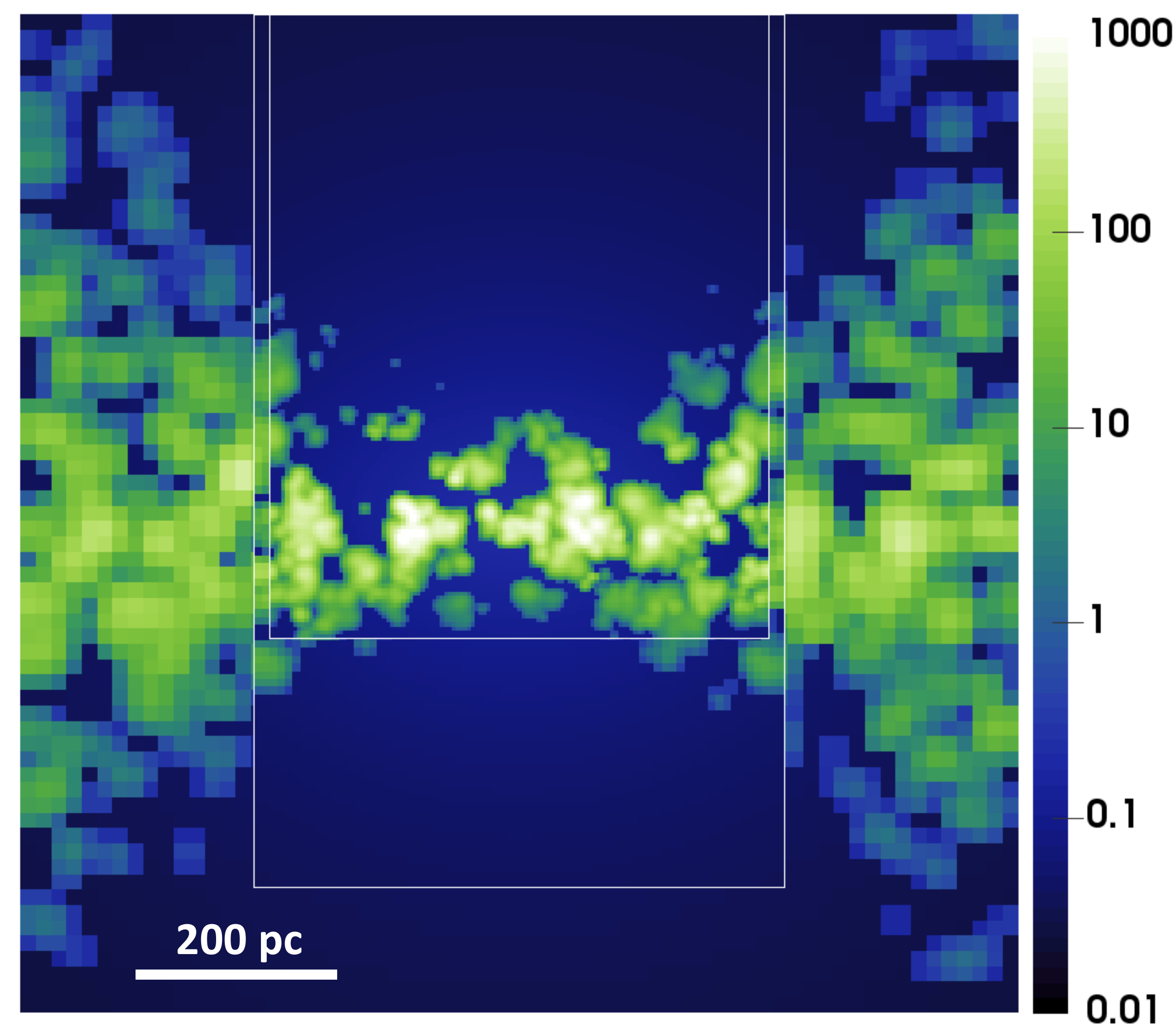}
\protect\caption[SMR Levels]{XZ plane slice of initial gas density (particles cm$^{-3}$) scaled logarithmically. White lines indicate SMR levels of refinement.}
\label{fig:test:smr} 
\end{figure}

In \citetalias{Tanner1} we employed a single grid, but in this paper we use the static mesh refinement (SMR) available in Athena \citep{Stone-Athena}.
We use two levels of refinement covering the starburst and wind region directly above the starburst extending to the +z boundary, as shown in Figure \ref{fig:test:smr}. 
The base grid is divided into $64^3$ cells, the first level divided into $64\times64\times112$ cells, and the second level divided into $128\times128\times160$ cells. 
This gives spatial resolution of $15.6$ pc on the base grid and $7.8$ and $3.9$ pc on each respective level. The resolution of the top level corresponds to our medium resolution models from \citetalias{Tanner1}. 

\subsection{Analytic Wind Velocity}\label{sec:setup:va}
As in \citetalias{Tanner1} we model the starburst as a spheroidal region where we inject thermal energy ($\dot{E}$) and mass ($\dot{M}$) at each time step. 
The analytical wind velocity depends on both $\dot{E}$ and $\dot{M}$ as,
\begin{equation}\label{eq:vA}
v_A = \sqrt{2\frac{\dot{E}}{\dot{M}}}.
\end{equation}
In this paper and in \citetalias{Tanner1} we refer to $v_A$ as the analytic wind velocity, whereas \citet{2005ARA&A..43..769V,2009ApJ...697.2030S} refer to it as the terminal wind velocity.
Both $\dot{E}$ and $\dot{M}$ are proportional to the total contribution from supernovae (SN) and stellar winds (SW) in the starburst. 
The total thermal energy injected into the ISM depends on the thermalization efficiency ($\epsilon$), which gives the fraction of SN and SW energy ($\dot{E}_{SN+SW}$) retained by the ISM as shown in Equation \ref{eq:Edot}. 
\begin{equation}\label{eq:Edot}
\dot{E} = \epsilon\dot{E}_{SN+SW}
\end{equation}
The total mass added to the ISM by the starburst includes contributions from SN and SW ($\dot{M}_{SN+SW}$) and cold gas ($\dot{M}_{cold}$) remaining from star formation that cannot be resolved by our simulations. 
The additional mass from cold gas can be factored in as a scaling factor ($\beta$) for $\dot{M}_{SN+SW}$ and is referred to as the mass loading factor \citep{2005ARA&A..43..769V,2009ApJ...697.2030S}, as shown in Equation \ref{eq:Mdot}.
\begin{equation}\label{eq:Mdot}
\dot{M} = \dot{M}_{SN+SW} + \dot{M}_{cold} = \beta\dot{M}_{SN+SW}
\end{equation}
We calculate $\dot{E}_{SN+SW}$ and $\dot{M}_{SN+SW}$ using Starburst99 models \citep{Leitherer1999}. We assume continuous star formation (CSF) with a \citet{2001MNRAS.322..231K} IMF, and Geneva stellar evolutionary tracks with solar metallicity. 
Both $\dot{E}_{SN+SW}$ and $\dot{M}_{SN+SW}$ scale with SFR \citep{2005ARA&A..43..769V} as,
\begin{eqnarray}
\dot{E}_{SN+SW} &=& 4.324e41~(\text{erg}~\text{s}^{-1})~\text{SFR}\label{eq:ESNSW}\\
\dot{M}_{SN+SW} &=& 0.1902~(\text{M}_\Sun~\text{yr}^{-1})~\text{SFR}.\label{eq:MSNSW}
\end{eqnarray}
With the above assumptions $\dot{E}_{SN+SW}$ and $\dot{M}_{SN+SW}$ are constant after $\sim5$ Myr. 
Thus we assume the starburst begins CSF 5 Myr before the start of our simulations, and we run our simulations for 1.5 Myr by which time a steady state wind has formed. 
Inserting Equations \ref{eq:Edot}, \ref{eq:Mdot}, \ref{eq:ESNSW}, and \ref{eq:MSNSW} into Equation \ref{eq:vA} and simplifying we get,
\begin{equation}\label{eq:vAsimp}
v_A = (1894~\text{km}~\text{s}^{-1})\sqrt{2\frac{\epsilon}{\beta}}.
\end{equation}
Thus $v_A$ does not depend on the SFR, but only $\epsilon$ and $\beta$. 
The actual velocity of the hot gas will be some fraction of $v_A$, and will be discussed in Section \ref{sec:res}. 
We use $v_A$ as a free parameter that we set in our models.

\subsection{Model Series}\label{sec:setup:mod}
We run two series of models, which we label series S and R, to test how the measured velocity of a galactic wind scales with the SFR and the SFR density (\sfrd) respectively. 
For all models we set $\epsilon=1.0$, $v_A$ to 1000, 1500 or 2000 \kms, and use Equation \ref{eq:vAsimp} to calculate $\beta$. 

The S series varies the SFR from 1 to 100 \msyr in steps of $0.1$ dex. 
For the S series we range over all SFRs with $v_A$ set to 1000, 1500 or 2000 \kms ($\beta = 7.1745$, 3.1887, 1.7936 respectively) for a total of 63 models. 
As in \citetalias{Tanner1}, our S series uses a starburst radius of 150 pc. 
Model numbers denote first the $v_A$ then the SFR. 
Thus model number S\_15\_79 has $v_A=1500$ \kms and SFR $=7.9$ \msyr.

The R series varies the radius (r) of the starburst from 50 to 500 pc in steps of $0.1$ dex. 
Each model in the R series has a fixed SFR of 10, 50 or 100 \msyr for a total of 33 models. 
The SFR density for this series is calculated using, \sfrd $= \text{SFR}/\pi r^2$. 
All R series models use $v_A=2000$ \kms ($\beta = 1.7936$). 
Model numbers denote first the SFR then the starburst radius. 
Thus model number R\_50\_79 has SFR 50 \msyr and starburst radius $79$ pc.

%Absorption Profiles
\section{Absorption Profiles}\label{sec:lines}
To probe the outflow velocities of different temperature regimes we generate synthetic absorption profiles for various silicon ions. 
We generate our absorption profiles by calculating the optical depth in each cell for a set of velocity channels directly above the starburst ($z>100$ pc) ranging from -2800 \kms to 200 \kms. 
We then integrate along a column of cells in the +z direction to get a total optical depth for each velocity channel, and then average each channel over all columns. 
We use a channel resolution of $\Delta v_{ch} = 0.25$ \kms. 

\begin{figure}[!ht]
\centering
\includegraphics[width=0.5\textwidth]{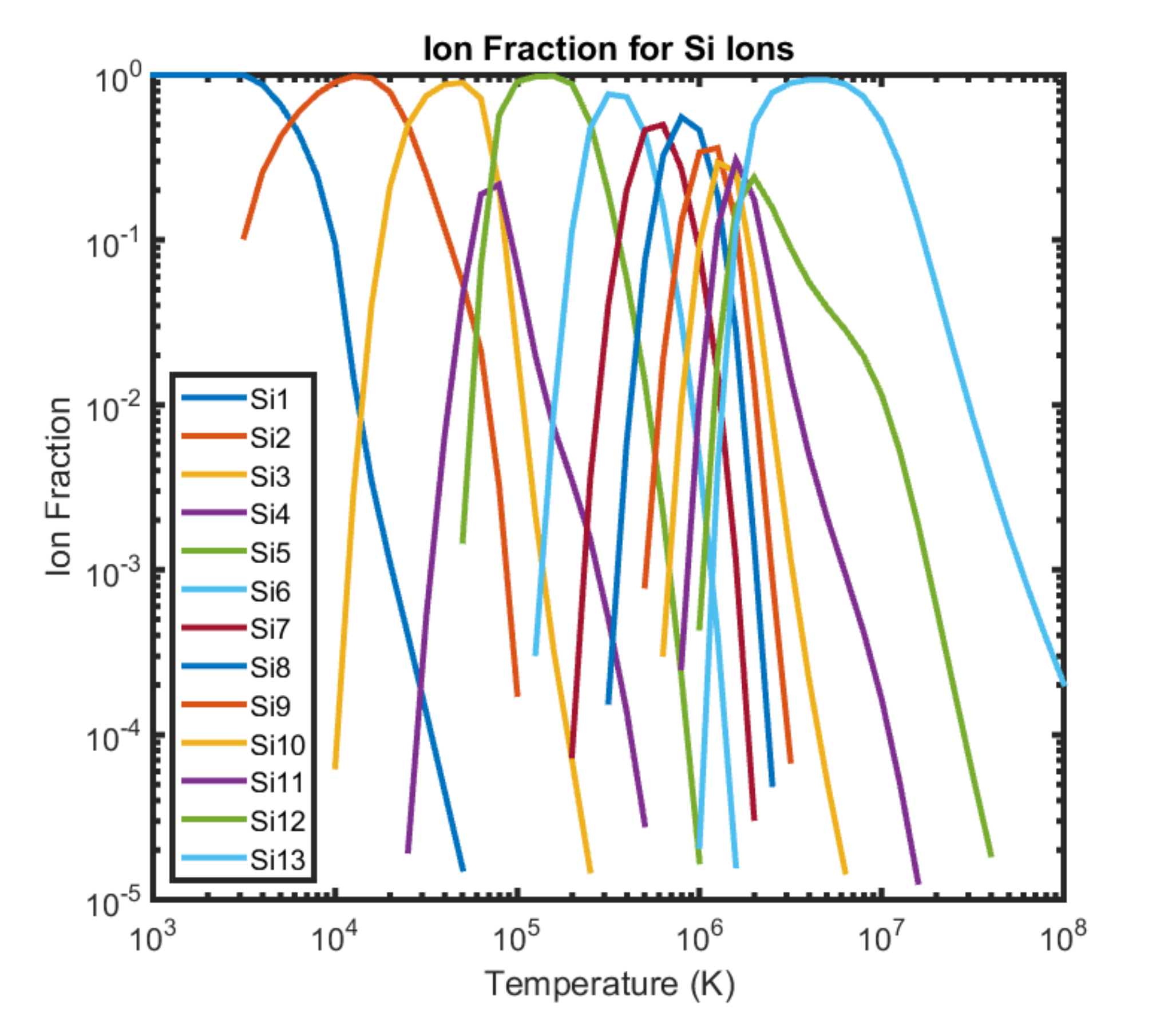}
\protect\caption[Ionization Fraction]{Ionization fractions for Silicon ions \citep{1998A&AS..133..403M} in collisional equilibrium at various temperatures.\label{fig:lines:ion}}
\end{figure}

The absorption coefficient for a single velocity channel ($v_{ch}$) is,
\begin{equation}\label{eq:lines:abs}
\kappa(v_{ch}) = N(v_{ch})a(v_{ch})
\end{equation}
where $N(v_{ch})$ is the column density and $a(v_{ch})$ is the absorption per atom. 
The column density is calculated from the cell density and the ionization fraction \citep{1998A&AS..133..403M}. 
For simplicity, and to compare with observations \citep{2015ApJ...811..149C,2016ApJ...822....9H}, we choose to use silicon ions for our analysis. 
The ionization fractions for Si I-XIII in collisional ionization equilibrium and at various temperatures are shown in Figure \ref{fig:lines:ion}. 

Contributions from Doppler broadening and spontaneous radiative transitions $a(v_{ch})$ is given as,
\begin{equation}\label{eq:lines:abspat}
a(v_{ch}) = \frac{\pi e^2}{m_e c}\frac{1}{\sqrt{\pi}}\frac{1}{\Delta \nu_{1/2}}f H(v_{ch}).
\end{equation}
Here $m_e$ is the mass of an electron, $c$ is the speed of light, $\Delta \nu_{1/2}$ is the half width half maximum (HWHM) of the Gaussian component, $f$ is the oscillator strength, and $H(v_{ch})$ is a Voigt profile.
The Gaussian HWHM is calculated \citep{kwok2007physics} using,
\begin{equation}
\Delta \nu_{1/2} = \frac{2}{c}\sqrt{\frac{2k_BT}{m}\ln(2)}\nu_0
\end{equation}
with $k_B$ Boltzmann's constant, $T$ the gas temperature, $m$ the atomic mass of the ion, and $\nu_0$ the frequency of the line center from the NIST Atomic Spectra database \citep{NIST_ASD}. 
We calculate the Voigt profile ($H(v_{ch})$) using Matlab code\footnote{\href{http://www.mathworks.com/matlabcentral/fileexchange/45058-deconvolution-mordenite-zeolite}{http://www.mathworks.com/matlabcentral/fileexchange/45058-deconvolution-mordenite-zeolite}} written by Dr. Nikolay Cherkasov which employs the method of \citet{Schreier20111010}.   

We calculate a normalized Voigt profile for each cell and then Doppler shift the profile using the $z$ velocity of the gas in the cell. 
The Doppler shift for each cell is calculated with respect to the systemic velocity of the galaxy, with negative velocities toward the observer and positive velocities away from the observer. 

\begin{figure}[ht]
\centering
\includegraphics[width=0.5\textwidth]{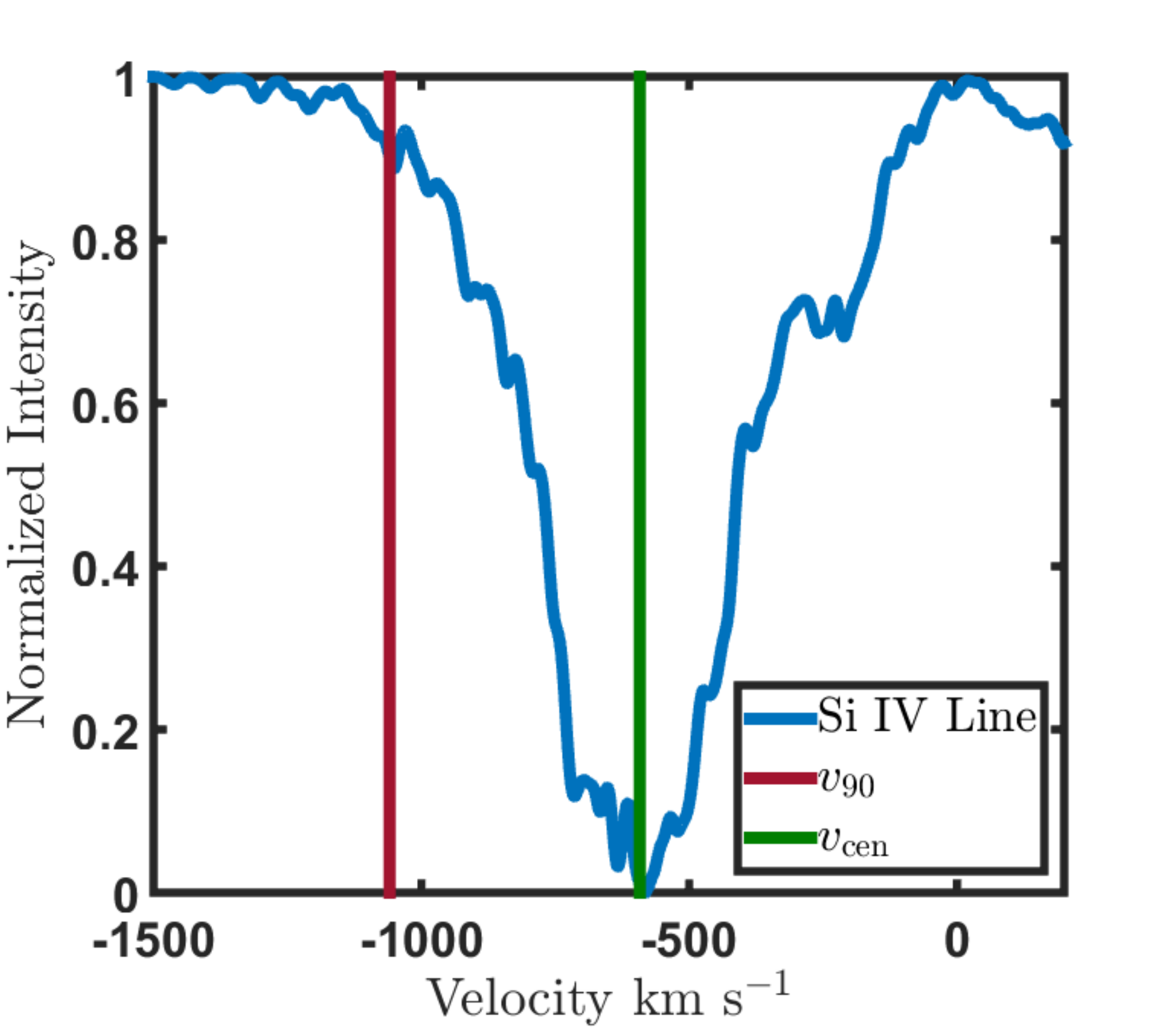}
\protect\caption{Synthetic absorption profile for a Si IV line. S\_20\_100 model with an analytic wind velocity of 2,000 \kms and a SFR of 10 \msyr. Vertical lines indicate $v_{cen}$ and $v_{90}$ velocities. \label{fig:lines:lineex}}
\end{figure}

Using Equation \ref{eq:lines:abs} we calculate the optical depth of each cell
\begin{equation}
\tau_i(v_{ch}) = \kappa_i(v_{ch}) dz.
\end{equation}
The optical depth for each velocity channel is then summed along a column in the z direction. The intensity for each velocity channel is calculated using,
\begin{equation}
I(v_{ch}) = I_0(v_{ch})\textrm{e}^{-\tau(v_{ch})}.
\end{equation}
The resulting profile is then averaged over all columns directly over the starburst and then re-normalized so that we can compare the profiles of different ions. 
Figure \ref{fig:lines:lineex} gives an example of a synthetic absorption profile for the Si IV line. 

Using the synthetic absorption lines, we measure outflow velocities of each ion in two ways.
$v_{cen}$ is the velocity at half of the full width at half maximum (FWHM). 
This effectively measures the average outflow velocity for the temperature range of the ion. 
$v_{90}$ is the velocity on the blueward side of the line where the absorption profile returns to 90\% of full intensity. 
This measures the maximum velocity of the gas for a specific temperature range. 
These velocity measures were adopted by \citep{2005ApJ...621..227M,2009ApJ...692..187W,2012ApJ...759...26E,2015ApJ...811..149C,2016MNRAS.457.1257H,2016ApJ...822....9H}, while others have used different, but comparable velocity measures \citep{2005ApJS..160..115R,2014ApJ...794..130B,2015ApJ...809..147H,2016A&A...588A..41C}. 
We use these velocities to determine the relationship between the galactic wind and the properties of the starburst. 
The vertical lines in Figure \ref{fig:lines:lineex} show $v_{cen}$ and $v_{90}$ for a synthetic absorption profile.

%Results
\section{Results}\label{sec:res}
In this section we investigate the relationships between $v_{cen}$ and $v_{90}$, and the analytic wind velocity ($v_A$ from Equation \ref{eq:vAsimp}), the SFR, and the SFR density ($\Sigma_{SFR}$). This allows us to determine how the outflow velocities of the multi-phase medium scale with different starburst properties. 

%Results: Ions
\subsection{Outflow Velocities of Different Ions}\label{sec:res:ions}
\begin{figure}[!ht]
\centering
\includegraphics[width=0.5\textwidth]{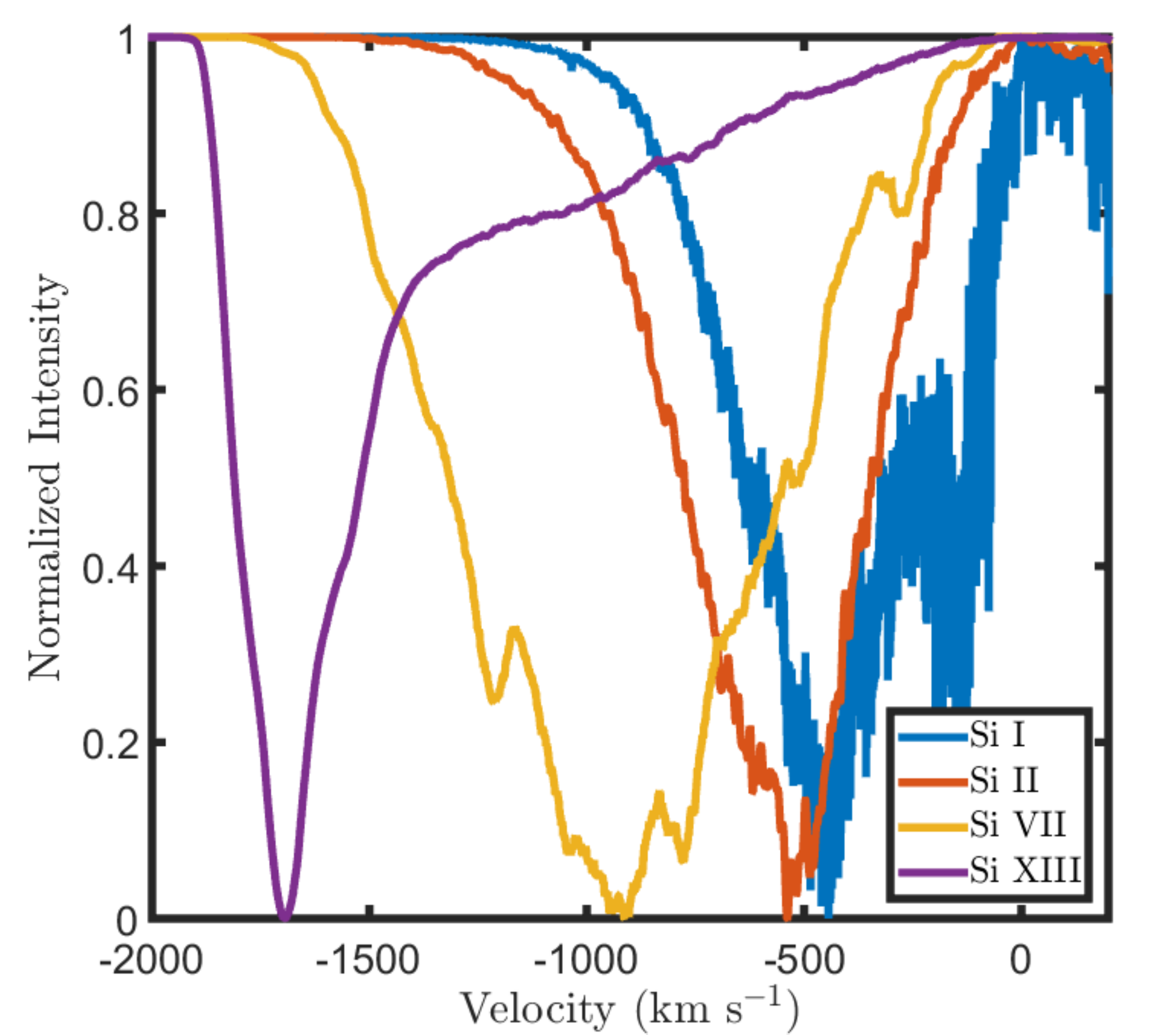}
\protect\caption[Si Absorption Lines]{Synthetic absorption lines for Si I, II, VII, and XIII from my S\_20\_1000 model, which has a $v_A$ of 2000 \kms and a SFR of 100 \msyr. \label{fig:lines:Siex}}
\end{figure}

In Figure \ref{fig:lines:Siex} we plot synthetic absorption lines for Si I, II, VII, and XIII for our S\_20\_1000 model. 
These four lines probe gas temperature ranges corresponding to $<1e4$ K, $1e4-2.5e4$ K, $4.5e5-7e5$ K, and $2e6-1e7$ K respectively. 
We note that hotter, highly ionized gas has a higher outflow velocity. 
This is in accordance with our results in \citetalias{Tanner1}. 
For the Si XIII line $v_{\rm cen} \approx 1700$ \kms, and $v_{90} \approx 1900$ \kms. 
These values are slightly less than the $v_A = 2000$ \kms of the S\_20\_1000 model. 
This difference is due to energy lost to cooling and kinetic energy transferred to the surrounding ISM. 
The cold gas, traced by Si I, moves much slower with $v_{\rm cen} \approx 300$ \kms, and $v_{90} \approx 800$ \kms. 
The $v_{\rm cen}$ and $v_{90}$ velocities for Si II are slightly higher than Si I, and the values for Si VII are midway between Si I and Si XIII. 

\begin{figure*}[h]
\centering
\includegraphics[width=1.0\textwidth]{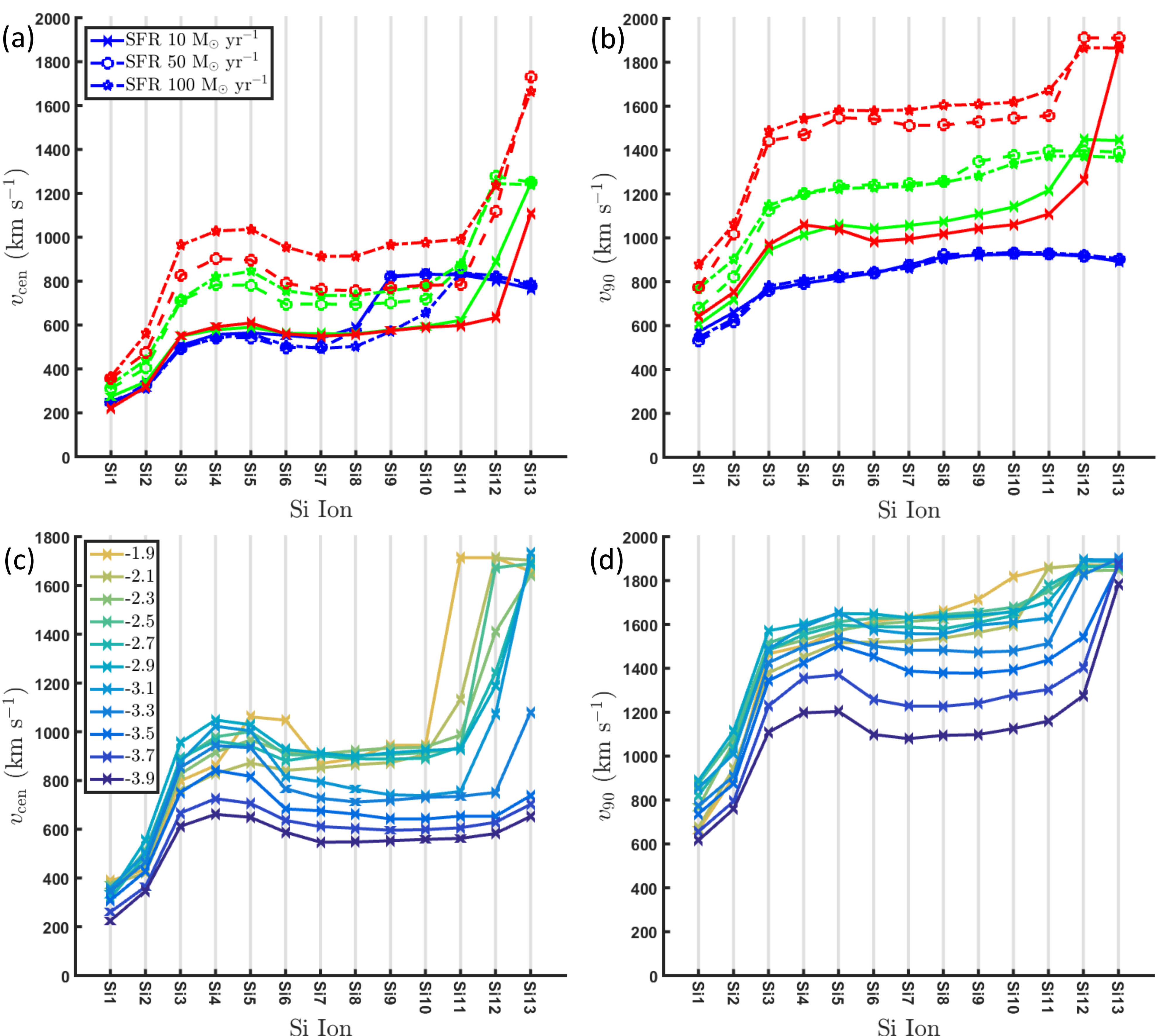}
\protect\caption{Outflow velocities of Si ions I-XIII for select S and R series models. Top, S series: Blue lines are for models with $v_A=1,000$ \kms, green for $v_A=1,500$, and red for $v_A=2,000$. Solid lines with 'x' indicate models with SFR of 10 \msyr, dashed lines with circles indicate a SFR of 50 \msyr, and dot dashed lines with pentagrams indicate models with SFR of 100 \msyr. (a) $v_{cen}$ velocity of all Si ions. (b) $v_{90}$ velocity of all Si ions. Bottom, R Series: Models have a SFR of 100 \msyr. Numbers in legend indicate $\log(\Sigma_{\rm SFR})$ \msyrpc of the models. (a) $v_{cen}$ velocity, (b) $v_{90}$ velocity.\label{fig:lines:SRions}}
\end{figure*}

To further determine how the velocity of the gas changes with increasing ionization we plot in Figure \ref{fig:lines:SRions} (Subplots (a) and (b)) $v_{cen}$ and $v_{90}$ respectively for Si I-XIII from our S series models. 
The plots include models with $v_A$ of $1000$, $1500$, and $2000$ \kms at SFR of 10, 50, and 100 \msyr. 

We find three distinct velocity regimes corresponding to Si I-II, Si III-XI, and Si XII-XIII, which correspond to temperatures ranges $<2.5e4$ K, $2.5e4-2e6$ K, and $>2e6$ K, respectively. 
The outflowing gas in the cold regime has much lower $v_{\rm cen}$ and $v_{90}$ than the warm regime, which in turn has lower velocities than the hot regime. 
In the warm regime the velocity plateaus and does not increase for several ions. 
For each ion, the measured velocity increases with increasing $v_A$, with a few notable exceptions. 
Models with $v_A=1000$ \kms show little variation in velocities for different SFRs, and the increase in velocity for hotter gas is less pronounced and occurs at Si VIII instead of Si XII or Si XIII as with models with higher $v_A$. 
The measured velocities for all ions from our model with a SFR 10 \msyr and $v_A=1500$ \kms are lower than the velocities from models with higher SFRs and $v_A=1500$ \kms. 
Our models with $v_A=2000$ \kms show a similar trend. 

In Figure \ref{fig:lines:SRions} (Subplots (c) and (d)) we plot the $v_{cen}$ and $v_{90}$ velocities respectively for Si I-XIII from our R series models. A similar three part grouping of velocities in cold, warm and hot regimes is evident. 
For each ion there is a trend of increasing velocity for increasing \sfrd. 

%Results: Scaling
\subsection{Scaling Relations}\label{sec:res:scale}
\subsubsection{Outflow Velocity vs. SFR}\label{sec:res:scale:S}
\begin{figure*}[h]
\centering
\includegraphics[width=1.0\textwidth]{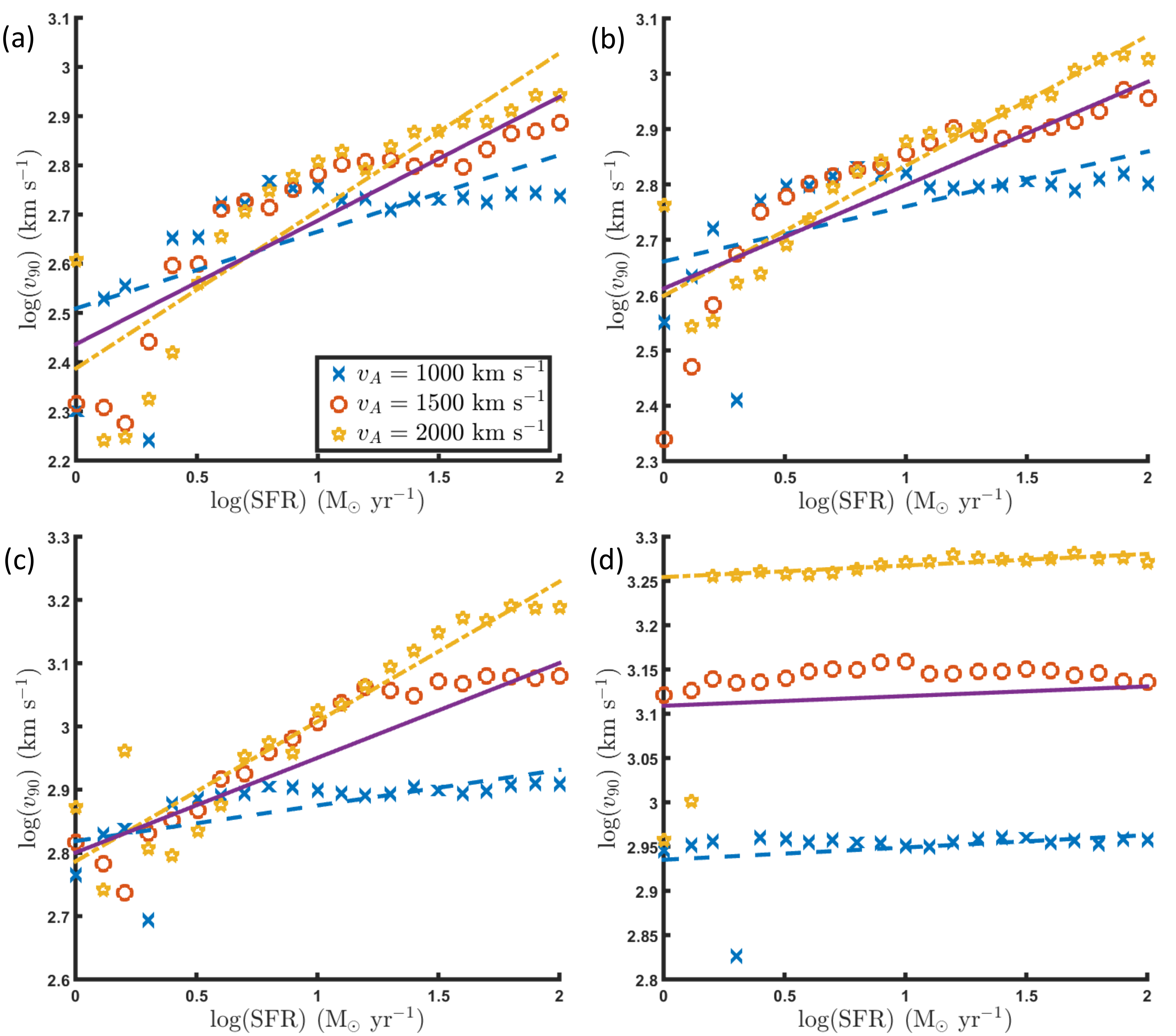}
\protect\caption{The $v_{90}$ velocity vs SFR for all S series models for select Si ions: (a) Si I, (b) Si II, (c) Si IV, (d) Si XIII. Lines indicate linear fits for models with $v_A=2000$ \kms (yellow, dot-dash line), $v_A=1000$ \kms (blue, dashed line), and all models together (purple, solid line). \label{fig:res:Sscale}}
\end{figure*}

\begin{figure*}[h]
\centering
\includegraphics[width=1.0\textwidth]{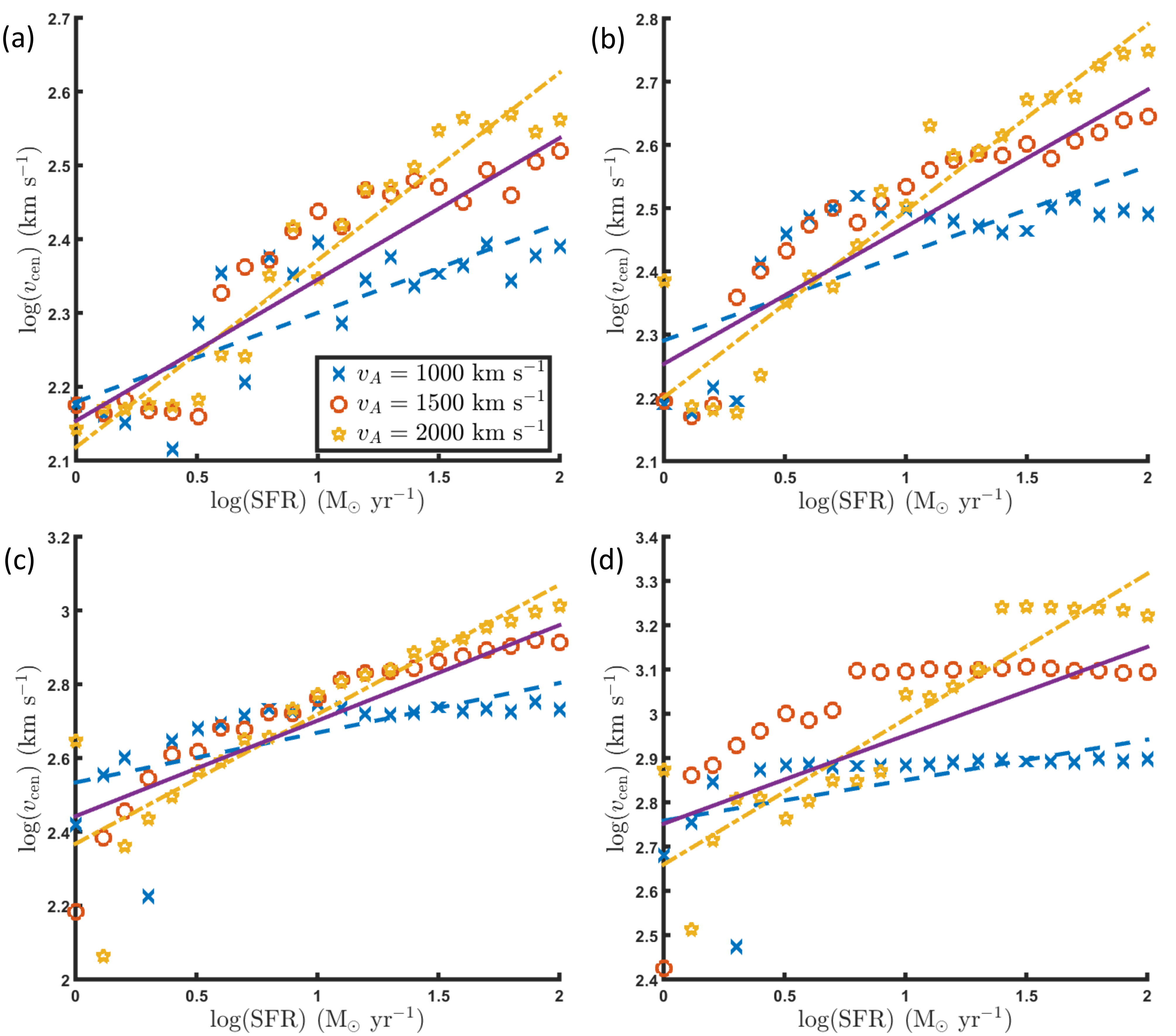}
\protect\caption{The $v_{\rm cen}$ velocity vs SFR for all S series models for select Si ions: (a) Si I, (b) Si II, (c) Si IV, (d) Si XIII. Lines indicate linear fits for models with $v_A=2000$ \kms (yellow, dot-dash line), $v_A=1000$ \kms (blue, dashed line), and all models together (purple, solid line). \label{fig:res:Sscalecen}}
\end{figure*}

To see how the maximum and central velocities scale with SFR, in Figures \ref{fig:res:Sscale} and \ref{fig:res:Sscalecen} we plot $v_{90}$ and $v_{\rm cen}$ respectively, versus SFR for Si I, II, IV and XIII from all S series models. 
The plots in Figure \ref{fig:res:Sscale} show that the maximum outflow velocity is correlated with the SFR, but only for some SFRs depending on the $v_A$ of the starburst. 
The measured $v_{90}$ velocity resulting from cold or warm gas correlates with the SFR until the velocity reaches some fraction of $v_A$, and then the relationship flattens out. 
This effect is most clearly seen in the plot for Si IV lines (Figure \ref{fig:res:Sscale} (c)). 
In the case of Si IV, the relationship flattens out when the $v_{90}$ velocity is $\sim 80$\% of $v_A$. 
A similar effect is observed for Si I and II $v_{90}$ velocities, with a turnover in the relationship at $\sim 50$\% and $\sim 65$\% of $v_A$ respectively.

The scaling relationship for Si XIII is flat, indicating that there is no correlation between SFR and outflow velocity for the hot gas. 
This is in agreement with Equation \ref{eq:vAsimp}, which shows that the analytic velocity does not depend on the the SFR. 
For hot gas, the $v_{90}$ velocity is roughly constant at $\sim 90$\% of $v_A$ for all SFRs. 

Figure \ref{fig:res:Sscalecen} shows that there are similar scaling relations between $v_{\rm cen}$ and SFR. 
But the relationship flattens out at $\sim 20$\%, $\sim 30$\% and $\sim 50$\% of $v_A$ for Si I, II and IV respectively. 
Interestingly, Si XIII now shows a correlation between SFR and the outflow velocity for low SFRs, with the relationship flattening out at $\sim 85$\% of $v_A$.

We perform a fit to each set of models with the same $v_A$ and then fit all models combined. 
The scaling relationship is of the form,
\begin{equation}
\log(v) = \delta\log(\text{SFR}) + \alpha
\end{equation}
For both $v_{90}$ and $v_{\rm cen}$ the scaling relation with SFR is steepest for models with a $v_A=2,000$ \kms and nearly flat for models with a $v_A=1,000$ \kms. 
For example, using $v_{\rm cen}$ velocities from Si IV lines, $\delta=0.35\pm0.08$ for models with $v_A=2,000$ \kms, and $\delta=0.14\pm0.08$ for models with $v_A=1,000$ \kms. 
For all data combined the fit falls between those two extremes with $\delta = 0.25\pm0.04$. 
All uncertainties are reported at a 95\% confidence level. 
The lower $\delta$ values are a result of including models with a SFR above the point where the relationship flattens out. 

As can be seen in Figures \ref{fig:res:Sscale} and \ref{fig:res:Sscalecen}, the same scaling relation exists for all models below the turnover point independent of $v_A$. 
We restrict our data set to models with SFRs below the turnover point and then fit the remaining data. 
The resulting slopes and y intercepts for these fits for $v_{90}$ and $v_{\rm cen}$ are given Table \ref{tab:fits}. 
As can be seen in Table \ref{tab:fits} for $v_{90}$ the slope is higher for low ionization and decreases with increasing ionization. 
But for $v_{\rm cen}$ the slope is lower for low ionization and increases with increasing ionization. 
Thus colder gas has a greater $\delta$ value than warmer gas for $v_{90}$, while the opposite is true for $v_{\rm cen}$. 

\begin{table}
\begin{center}
\caption{\label{tab:fits}Fit data for outflow velocities vs. SFR from S series models. Only models below the turnover point are used.}
\begin{tabular}{lll} \toprule
Ion & Slope ($\delta$) & Intercept ($\alpha$) \\
\midrule
\multicolumn{3}{l}{$v_{90}$ Velocities}\\
\midrule
Si I  & $0.38 \pm 0.08$ & $2.37 \pm 0.06$ \\
Si II & $0.28 \pm 0.06$ & $2.56 \pm 0.05$ \\
Si III & $0.26 \pm 0.03$ & $2.72 \pm 0.03$ \\
Si IV & $0.25 \pm 0.03$ & $2.76 \pm 0.03$ \\
\midrule
\multicolumn{3}{l}{$v_{\rm cen}$ Velocities}\\
\midrule
Si I  & $0.28 \pm 0.03$ & $2.11 \pm 0.03$ \\
Si II & $0.32 \pm 0.05$ & $2.20 \pm 0.04$ \\
Si III & $0.34 \pm0.06$ & $2.38 \pm 0.05$ \\
Si IV & $0.37 \pm 0.08$ & $2.39 \pm 0.06$ \\
\bottomrule
\end{tabular}
\end{center}
\end{table}

%Results: Scaling: SFR density
\subsubsection{Outflow Velocity vs. SFR Density}\label{sec:res:scale:R}
\begin{figure*}[h]
\centering
\includegraphics[width=1.0\textwidth]{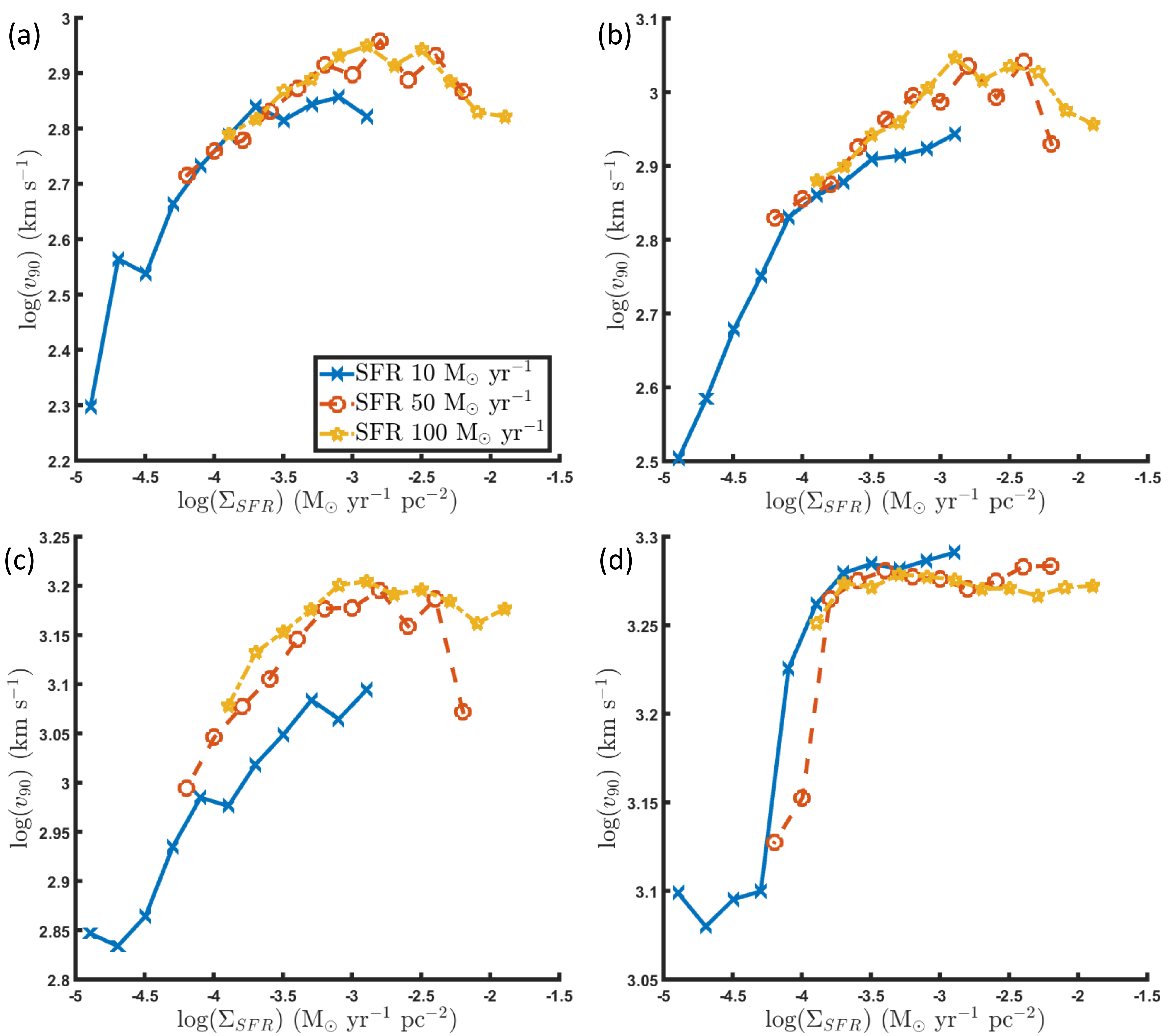}
\protect\caption{The $v_{90}$ velocity vs \sfrd for all R series models for select Si ions: (a) Si I, (b) Si II, (c) Si IV, (d) Si XIII. \label{fig:res:Rscale}}
\end{figure*}

Using our R series models we determine the relationship between the outflow velocity and the SFR density (\sfrd). 
In Figure \ref{fig:res:Rscale} we plot the $v_{90}$ velocity versus \sfrd for all R series models. 
For our R series we set $v_A = 2000$ \kms and tested three SFRs (10, 50 and 100 \msyr) while varying the size of the starburst to achieve a range of \sfrd. 

As can be seen in Figure \ref{fig:res:Rscale}, much like with Figure \ref{fig:res:Sscale}, the outflow velocity is correlated with \sfrd, up to a point, and then the correlation flattens out. 
For all ions, including Si XIII, the turnover point for $v_{90}$ is at $\log(\Sigma_{\rm SFR})\approx -3.5$ (M$_{\odot}$ yr$^{-1}$ pc$^{-2}$). 
The turnover point for $v_{\rm cen}$ is slightly higher at $\log(\Sigma_{\rm SFR})\approx -3.0$ (M$_{\odot}$ yr$^{-1}$ pc$^{-2}$). 
Because we set $v_A = 2000$ \kms for our R series these turnover points are higher than they would be for models with a lower $v_A$. 
As shown in Section \ref{sec:res:scale:S}, models with a lower $v_A$ have a lower turnover point. 
If we compare the velocities at the turnover point for each ion in our R series we find that they are similar to the velocities at the turnover point for our S series models with $v_A = 2000$ \kms. 

We perform a similar analysis as was done in Section \ref{sec:res:scale:S} on our R series and determine scaling relations for models below the turnover point. 
The fit parameters are summed up in Table \ref{tab:fits:R}. 
Similar to our results for our S series, the slopes calculated using $v_{90}$ velocities are steeper for colder gas and shallower for warmer gas. 
But the opposite trend between cold and warm gas slopes calculated using $v_{\rm cen}$ velocities is not as clear, since Si I produces the steepest slope. 

It is interesting to note that the SFR does not affect the outflow velocity, even below the turnover point. In Figure \ref{fig:res:Rscale} for each ion we plot three sets of models with different SFRs. 
There is significant overlap among the three sets, and all have similar scaling relations, turnover points, and velocities above the turnover point. 

\begin{table}
\begin{center}
\caption{\label{tab:fits:R}Fit data for outflow velocities vs. \sfrd from R series models. Only models below the turnover point are used.}
\begin{tabular}{lll} \toprule
Ion & Slope ($\delta$) & Intercept ($\alpha$) \\
\midrule
\multicolumn{3}{l}{$v_{90}$ Velocities}\\
\midrule
Si I  & $0.32 \pm 0.07$ & $4.0 \pm 0.3$ \\
Si II & $0.28 \pm 0.05$ & $3.9 \pm 0.2$ \\
Si III & $0.25 \pm 0.05$ & $4.0 \pm 0.2$ \\
Si IV & $0.22 \pm 0.05$ & $3.9 \pm 0.2$ \\
\midrule
\multicolumn{3}{l}{$v_{\rm cen}$ Velocities}\\
\midrule
Si I  & $0.35 \pm 0.08$ & $3.7 \pm 0.3$ \\
Si II & $0.26 \pm 0.04$ & $3.5 \pm 0.1$ \\
Si III & $0.32 \pm 0.07$ & $3.9 \pm 0.3$ \\
Si IV & $0.32 \pm 0.07$ & $4.0 \pm 0.3$ \\
\bottomrule
\end{tabular}
\end{center}
\end{table}

%Discussion
\section{Discussion}\label{sec:dis}
\subsection{Absorption Line Shape}\label{sec:dis:shape}
There are two important characteristics of our synthetic absorption profiles, the smoothness and the asymmetry of the lines. 
In Figure \ref{fig:lines:Siex} the Si I line is not smooth, but has a large number of jagged spikes especially at lower velocities. 
As explained in \citetalias{Tanner1}, the colder gas entrained in the wind is confined in filamentary structures with lower velocity than the surrounding hot gas. 
As these filaments fragment in the wind due to Kelvin-Helmholtz instabilities they produce many dense cores \citep{CooperII}. 
The contributions from these dense cores embedded in the filaments produce the jagged shape of the Si I line. 
The smoothness of the absorption line, or lack thereof, reflects the distribution of vertical velocities of the dense cores and filaments. 
We note that there is measurable absorption at positive velocities from dense cores that have been elevated above the galactic disk but whose vertical movement has stalled. 

The Si II line is significantly smoother than the Si I line. 
This indicate that the gas traced by the Si II line is significantly less clumpy than the colder gas traced by Si I. 
As the dense cores embedded in the wind are disrupted, the cold gas is ablated and accelerated to a higher velocity while being heated by the wind. 
This produces the asymmetric profiles of the Si I and II lines. 
These asymmetries have been observed in absorption lines from several galaxies \citep{2009ApJS..181..272G,2012ApJ...751...51J,2013ApJ...765..118W,2015ApJ...809..147H,2015ApJ...810..104A,2015ApJ...811..149C,2016MNRAS.457.3133C}. 
The asymmetric tail to higher velocities results from ablated gas being accelerated as it is entrained in the hot wind. 
The Si IV (Figure \ref{fig:lines:lineex}) and Si VII (Figure \ref{fig:lines:Siex}) lines are not highly asymmetric due to contributions from both slower gas ablated off of the cooler dense cores, and warmer gas beginning to mix with the hot gas that fills the superbubble created by the starburst. 
The Si VII line is not entirely smooth as it traces the large scale structure of filaments embedded in the wind. 
The asymmetry of the Si XIII line skews in the opposite direction. The long tail to lower velocities results from the hot gas being accelerated as it moves off the galactic disk. 
It reaches terminal velocity at $\sim 300$ pc above the galactic center to form the deep dip in absorption in Figure \ref{fig:lines:Siex}. 

%Discussion: Outflow Velocities
\subsection{Outflow Velocities}\label{sec:dis:vel}
As shown in Figures \ref{fig:res:Sscale} and \ref{fig:res:Sscalecen} the measured outflow velocity for neutral and slightly ionized gas increases with SFR. 
Considering just Si II (Subplot (b) in both Figures \ref{fig:res:Sscale} and \ref{fig:res:Sscalecen}), at SFR $= 1$~\msyr, $v_{\rm cen}\approx 160$~\kms and $v_{90}\approx 400$~\kms. 
At SFR $= 100$~\msyr there is significant divergence in the measured velocities depending on the $v_A$, but the highest measured velocities are $v_{\rm cen}\approx 560$~\kms and $v_{90}\approx 1070$~\kms.

\citet{2015ApJ...811..149C} use Si II lines to measure outflow velocities from their sample of star forming galaxies with SFRs ranging from 0.01 \msyr to 100 \msyr. 
At SFR $= 1$~\msyr they measure $v_{\rm cen}\approx 50$~to 150 \kms and $v_{90}\approx 350$~\kms. 
At higher SFRs they also have increased scatter in their measured velocities as we have in our data.
At SFR $\lesssim 100$~\msyr their highest velocities are $v_{\rm cen}\approx 500$~\kms and $v_{90}\approx 1100$~\kms. 
Thus our measured velocities correspond to the outflow velocities found by \citet{2015ApJ...811..149C}. 
We find similar agreement with the measured velocities from \citet{2015ApJ...809..147H} and \citet{2016ApJ...822....9H}, but only for our models with the highest $v_A$. 

The sample of galaxies used by \citet{2015ApJ...809..147H} and \citet{2016ApJ...822....9H} overlap with the sample used by \citet{2015ApJ...811..149C}, as both use data from \citet{2011ApJ...730....5H} and \citet{2015ApJ...810..104A}, but the overlap only accounts for about a third to a half of the galaxies. 
This results in increased scatter in the measured velocities at higher SFRs for \citet{2015ApJ...809..147H}.

As shown in Figure \ref{fig:lines:SRions} the velocities of neutral or slightly ionized gas are significantly lower than the velocity of the hot, highly ionized gas. 
Higher outflow velocities for higher ionization have been found in some surveys of starburst galaxies \citep{2009ApJS..181..272G,2016A&A...588A..41C}. 
Thus surveys which rely on absorption lines which trace neutral gas \citep{2000ApJS..129..493H,2005ApJ...621..227M}, or low ionized gas \citep{2014ApJ...794..156R,2015ApJ...811..149C,2015ApJ...809..147H,2016ApJ...822....9H,2016A&A...588A..41C} will underestimate outflow energetics. 

The difference is more pronounced at low SFRs below the turnover point where there is degeneracy in the measured velocities for models with differing $v_A$. 
This difference is greatest for models with high $v_A$. 
The degeneracy below the turnover point means the degree to which the energy of the outflow is underestimated cannot be known. 
At higher SFRs above the turnover point, the velocity of neutral and low ionized gas can more reliably be used as a proxy for the outflow velocity of the hot gas. 

%Discussion: Scaling Relations
\subsection{Scaling Relations}\label{sec:dis:scale}
A scaling relation between outflow velocity and SFR has been found in numerous surveys of starburst galaxies \citep{2000ApJS..129..493H,2005ApJ...621..227M,2005ApJS..160..115R,2009ApJ...692..187W,2012ApJ...760..127M,2012ApJ...759...26E,2012ApJ...758..135K,2014ApJ...794..130B,2014ApJ...794..156R,2015ApJ...811..149C,2016MNRAS.457.3133C,2015ApJ...809..147H,2016ApJ...822....9H,2016A&A...588A..41C}. 
A similar relationship has been found between the outflow velocity and \sfrd \citep{2010AJ....140..445C,2012ApJ...758..135K,2014ApJ...794..156R,2015ApJ...811..149C,2016MNRAS.457.3133C,2015ApJ...809..147H,2016ApJ...822....9H}. 

All of these surveys rely on either optical or UV absorption lines, which trace warm, not hot gas. 
Our results in Section \ref{sec:res:scale:S} show that only the hot gas follows the relation given in Equation \ref{eq:vAsimp}, where the outflow velocity does not depend on the SFR. 
The outflow velocity of the warm gas does depend on the SFR, but only if the velocity is below some fraction of $v_A$. 
The turnover point where the relationship flattens out, also called the saturation point \citep{2015ApJ...809..147H,2016ApJ...822....9H}, depends on the $v_A$ associated with the starburst and the ion  being used. 
Below the turnover point the scaling relation for each ion is uniform and does not depend on $v_A$. 

Many surveys find a correlation between outflow velocity and SFR, but there is no agreement about the slope of the scaling relation ($\delta$). 
For example, using $v_{90}$ velocities \citet{2005ApJS..160..115R} find $\delta = 0.21$, \citet{2009ApJ...692..187W} find $\delta = 0.38$, \citet{2015ApJ...811..149C} find $\delta = 0.081$, and \citet{2016ApJ...822....9H} find $\delta = 0.32$. 
Yet using $v_{\rm cen}$ velocities \citet{2005ApJ...621..227M} find $\delta = 0.35$, and \citet{2015ApJ...811..149C} find $\delta = 0.22$. 

Our results show that the lower $\delta$ values result from samples comprising a combination of galaxies with high and low $v_A$ velocities, and therefore a mix of galaxies above and below the turnover point. 
Including a mix of galaxies in the sample would produce increased scatter in the data at higher SFRs or \sfrd because it would include galaxies above the turnover point in the relationship.  
In contrast, the higher $\delta$ values would come from samples of galaxies with either relatively uniform, and high, $v_A$, or a sample restricted to the lower range of SFRs below the turnover point. 

For example, \citet{2015ApJ...809..147H} calculate $\beta\approx 2$ for all of the galaxies in their sample. 
This would imply that the galaxies in their sample have relatively uniform $v_A$ velocities between 1500 and 2000 \kms, putting the turnover point in the velocity vs. SFR relation high enough that it would not significantly lower their $\delta$. 
When we compare the galaxy samples used we find that high $\delta$ values come from samples consisting of ULIGs \citep{2005ApJ...621..227M}, high red-shift, luminous galaxies \citep{2009ApJ...692..187W}, or UV-bright galaxies \citet{2016ApJ...822....9H}. 
Lower $\delta$ values come from IR-bright galaxies \citep{2005ApJS..160..115R} or samples of local starburst galaxies with a variety of morphologies \citep{2015ApJ...811..149C}. 

While \citet{2015ApJ...809..147H} do not find a turnover in their relation between velocity and SFR, they note a turnover in the relation between velocity and \sfrd at $\sim 10^{-5}$ \msyrpc. 
In \citet{2016ApJ...822....9H} they find the turnover point to be $\sim 10^{-4}$ \msyrpc. 
The latter value is similar to our turnover point of $\sim 10^{-3.5}$ \msyrpc from our R series, which has a $v_A=2000$ \kms. 
We would expect their turnover point to be slightly lower because the mass loading factor of their galaxies would imply a lower $v_A$ than our R series models. 
Also, we set the radius of our starburst, and from that calculate \sfrd, while they find the starburst radius using the half light radius in the UV. 
So we would expect affect the measured turnover point in the relationship. 

\section{Conclusions}\label{sec:end}
Using synthetic absorption lines generated from our series of 3D starburst models, we have investigated how the velocity of starburst-driven galactic winds scales with both SFR and \sfrd. 
As implied by Equation \ref{eq:vAsimp} the terminal wind velocity should not depend on SFR, and by extension \sfrd. 
Our simulations, both in this paper and in \citetalias{Tanner1}, have shown that the gas entrained in a starburst-driven wind does not move as a single coherent structure. 
Cold gas moves much slower than warm gas, which in turn moves slower than hot, highly ionized gas. 
But the relative velocities of these three gas phases do not relate to each other in a simple straight forward manner. 

The measured velocity of the hot, highly ionized, gas is independent of SFR, and instead depends on $v_A$, and by extension, the thermalization efficiency and mass loading factor associated with the starburst. 
Meanwhile both the warm and cold gas scale with SFR, but only up to a point. 
The point where the scaling relation flattens out and the velocity becomes independent of SFR is determined again by $v_A$. 
Below this point the scaling relation is independent of $v_A$, and depends entirely on what temperature regime is being probed. 
Because of this, there is degeneracy in the velocity for different values of $v_A$, which means at low SFRs neutral or warm gas tracers cannot be used as proxies for the velocity of the hot, highly ionized gas. 
Using neutral or low ionized gas to determine outflow energetics will underestimate the total energy by an indeterminate amount. 
Our models show that at low SFRs it is possible to have a high velocity ($\sim 2000$~\kms) outflow while the neutral gas embedded in the wind would have a much lower velocity ($\sim 300$~\kms). 
 
The relationship between the outflow velocity and \sfrd behaves in a similar way. 
There is a point, determined by $v_A$, where the relationship flattens out, but below that point the velocity scales with \sfrd. 
The turnover has been found \citep{2015ApJ...809..147H}, and is in agreement with our results. 
Even though we did not address how outflow velocity scales with other galaxy characteristics such as total stellar mass and the specific SFR, similar effects in the scaling relations may be present. 

Our results indicate that the differences in the measured scaling relations stem from sample selection criteria that group galaxies with either a mix of $v_A$ velocities or uniform velocities. 
This would explain why \citet{2014ApJ...794..156R} found no correlation between the outflow velocity and either SFR and \sfrd, while \citet{2015ApJ...811..149C} found a correlation for both, but with a modest slope. 
But \citet{2015ApJ...809..147H} found a steeper slope to the correlation for both SFR and \sfrd. 
If we only use models with a high $v_A$ we reproduce the high slopes found by \citet{2015ApJ...809..147H}, and when we include models a mix of both high and low $v_A$ we reproduce the slopes found by \citet{2015ApJ...811..149C}. 
If we were to extend our models to include higher mass loading factors, and therefore lower $v_A$, our models would show no correlation, with a high scatter, as found by \citet{2014ApJ...794..156R}. 
Thus when considering the scaling relations between outflow velocity, SFR and \sfrd we cannot treat a sample of star forming galaxies as a single coherent group. 

\acknowledgments{}
NASA Herschel grant NHSC-OT-1-1436036 and NC Space Grant supported this work.

\medskip
\bibliography{./generalbib}
\end{document}